\documentclass[prb,twocolumn]{revtex4}

\usepackage{graphics}

\newcommand{\be}{\begin{equation}}
\newcommand{\ee}{\end{equation}}
\newcommand{\bea}{\begin{eqnarray}}
\newcommand{\eea}{\end{eqnarray}}

\begin{document}


\title{Global Phase Diagram of $\nu = 2$ Quantum Hall Bilayers in
Tilted Magnetic Field}
\author{Anna Lopatnikova$^{1*}$}
\author{Steven H. Simon$^{2}$}
\author{Eugene Demler$^{1}$}
\affiliation{
$^{1}$Department of Physics, Harvard University, Cambridge, MA 02138 \\
$^{2}$Lucent Technologies Bell Labs, Murray Hill, NJ 07974 }

\date{\today}

\begin{abstract}

We consider a bilayer quantum Hall system at total filling
fraction $\nu=2$ in tilted magnetic field allowing for charge
imbalance as well as tunneling between the two layers. Using an
``unrestricted Hartree Fock,'' previously discussed by Burkov and
MacDonald (Phys Rev B {\bf 66} 115323 2002), we examine the zero
temperature global phase diagrams that would be accessed
experimentally by changing the in-plane field and the bias voltage
between the layers while keeping the tunneling between the two
layers fixed. In accordance with previous work, we find symmetric
and ferromagnetic phases as well as a first order transition
between two canted phases with spontaneously broken $U(1)$
symmetry. We find that these two canted phases are topologically
connected in the phase diagram and, reminiscent of a first order
liquid-gas transition,  the first order transition line between
these two phases ends in a quantum critical point. We develop a
physical picture of these two phases and describe in detail the
physics of the transition.

\end{abstract}

\maketitle

\section{Introduction}
\label{sec:intro3}

Over the last fifteen years, one of the most exciting frontiers in
two dimensional electron physics has been the study of quantum
Hall bilayers\cite{dassarma}. Stacking two quantum Hall systems a
small distance away from each other serves two main purposes:
First of all, it is a way of creating a multicomponent system in
which the layer index of the electron plays the role of an
isospin. Secondly, it is a step in adding another dimension to the
traditionally two dimensional quantum Hall medium.

One of the most striking phenomena that illustrates the richness
of the physics of bilayer quantum Hall systems is the
commensurate-incommensurate transition observed in systems with
filling fraction $\nu=1$ subjected to tilted magnetic
fields\cite{dassarma,wen:92,murphy:94,yang:94,moon:95,yang:96}.
For typical parameters in these systems, the spin degrees of
freedom are frozen out and the only important discrete degree of
freedom is the layer index (isospin). Thus only two processes are
important:  interlayer tunneling and Coulomb interactions.   When
such a bilayer system is subjected to tilted magnetic fields, the
electrons tunneling between the layers enclose flux quanta. In the
presence of a sufficiently strong in-plane magnetic field it may
be favorable for the system to forgo tunneling in order to escape
destructive interference induced by the in-plane field. This
creates an opportunity for a phase transition to occur between a
phase in which tunneling plays the central role and a phase in
which tunneling is effectively zero.

Recently, it has been shown\cite{burkov:02} that a somewhat
different phase transition can occur in bilayer systems with a
total filling fraction $\nu=2$.  The $\nu=2$ bilayer system is
enriched not only by the presence of two electrons per flux
quantum, but most importantly by the role of real spin.  In
contrast to $\nu=1$ bilayers where the real spin degrees of
freedom are frozen out, in $\nu=2$ bilayers, spin degrees of
freedom are {\it entangled} with the isospin degrees of freedom as
a result of Fermi statistics (Pauli exclusion principle, in other
words). Thus, the $\nu=2$ bilayers exhibit a richer phase diagram
even in perpendicular field, and, as we will see later, an even
more intriguing phase diagram in tilted field.

In 1997, Zheng, Radtke, and Das Sarma (ZRD)
\cite{zheng:97,dassarma:98} predicted that the bilayer quantum
Hall systems at total filling fraction $\nu=2$ can exhibit a novel
spontaneously symmetry-broken phase.  ZRD performed a
time-dependent Hartree-Fock study of spin-density excitations of
the $\nu=2$ bilayers and found that, under experimentally
attainable conditions (at finite Zeeman energy and interlayer
tunneling), the spin-density mode softens, signaling a phase
transition to a new, symmetry-broken, quantum Hall state.  The
novel state was found to have a finite magnetization, like the
simple ferromagnetic state, ($F$), that occurs in the limit of
large Zeeman energy;  it also exhibited interlayer phase
coherence, like the spin-singlet state, ($S$), stabilized by
strong interlayer tunneling.  In addition, the new state was found
to possess antiferromagnetic correlations.  ZRD dubbed the novel
state ``canted'', since schematically the state can be viewed as a
one in which the spins in the opposite layers are canted away from
the magnetic field in opposite directions.  The canted ground
state breaks the $U(1)$ symmetry associated with rotations around
the direction of the magnetic field; this spontaneously broken
symmetry is behind the formation of the soft (Goldstone)
spin-density mode found by ZRD.

The canted state is a pure many-body state, stabilized by Coulomb
interactions.  In the absence of the interactions, there would be
a first-order phase transition between the ferromagnetic phase and
the spin-singlet state.  Coulomb interactions effectively mix the
ferromagnetic state and the spin-singlet state around the would-be
first-order phase transition (when the energy splitting between
these states is small), giving rise to the canted phase.  Because
Coulomb interactions somewhat favor the ferromagnetic state, a
finite amount of tunneling is crucial for the stability of the
canted phase.

Brey, Demler, and Das Sarma (BDD) \cite{brey:99}, however, showed,
that the canted phase can be extended to the region of finite
Zeeman energy and infinitely small tunneling by creating a charge
imbalance between the layers.  The charge imbalance can be induced
by an external bias voltage applied perpendicularly to the system.
In the charge-unbalanced regime, another many-body phase with a
spontaneously broken U(1) symmetry, but with vanishing
antiferromagnetic order parameter, was discovered at zero
tunneling.  This phase --- the $I$-phase --- is continuously
(without a phase transition) connected to the antiferromangetic
canted phase.  MacDonald, Rajaraman, and Jungwirth (MRJ)
\cite{macdonald:99} pointed out in their thorough Hartree-Fock
study of the $\nu=2$ bilayer phase diagram that the $I$-phase is
akin to the spontaneous interlayer phase coherent phase that
occurs in the $\nu=1$ bilayers in the absence of tunneling
\cite{wen:92}.  MRJ therefore suggested that the spontaneous
interlayer phase coherence of the $I$-phase may lead to
interesting effects in tilted fields, closely related to the
commensurate-incommensurate transition of the $\nu=1$ bilayers
\cite{murphy:94,yang:94,moon:95,yang:96}.

In a recent paper, Burkov and MacDonald (BM) \cite{burkov:02}
explore this possibility.  Indeed they find that a tilted field
applied to a charge-unbalanced $\nu=2$ bilayer system induces a
new quantum phase transition within the canted phase.  Their
``unrestricted Hartree-Fock'' analysis shows that the phase
transition is first-order and is between two commensurate phases:
At low tilt angles the phase is a simple canted commensurate
phase, in which the layer degree of freedom (isospin) is
commensurate with the in-plane component of the magnetic field. At
higher tilt angles a phase transition occurs to a phase, in which
not only the isospin, but also the {\it spin} becomes commensurate
with the in-plane field (Fig.~\ref{fig:qs}). Thus, BM conclude,
the phase transition is not a commensurate-incommensurate
transition.

In this paper, their results are extended and given a physical
explanation.  Using the ``unrestricted Hartree-Fock''
approximation, we find that the new first-order phase transition
{\it terminates with a quantum critical transition embedded within
the canted phase} (Fig.~\ref{fig:tilt}).   The simple commensurate
and the spin-isospin commensurate canted phases are continuously
connected to each other ---  much like the familiar example of
water and its vapor, the two phases possess the same symmetry
properties (the spontaneously broken $U(1)$ in our case). To
illustrate the novel first-order transition and the quantum
critical transition that terminates it, we present a series of
phase diagrams (Fig.~\ref{fig:tilt}).  Each phase diagram is
obtained for fixed perpendicular-field Zeeman energy and tunneling
strength, and the axes on the phase diagrams are the in-plane
field and the bias voltage.  We find this choice of axes
particularly suitable, since current experimental techniques allow
to vary the bias voltage and the in-plane field {\it in situ} over
a wide range of values.  All the phases on a given phase diagram
can therefore be accessed on a single sample;  this should
facilitate the detection of the novel phase transitions and the
new phases.  In an upcoming publication \cite{lopatnikova:03b}, we
obtain the collective modes in various parts of the phase diagram,
thus providing signatures of different phases and phase
transitions for possible light-scattering experiments.

In addition to our new Hartree-Fock results, we present a detailed
physical discussion of the surprising behavior of the $\nu=2$
bilayers in tilted magnetic field.  We show that the exotic
spin-isospin commensurate canted phase is closely related to the
$I$-phase, as illustrated by a comparison of order parameters in
the two phases (Fig.~\ref{fig:op}).  In fact, the spin-isospin
commensurate phase rapidly converges to the $I$-phase as the tilt
angle is increased.  We use the symmetry properties of the
$I$-state to give a simple explanation to the spin-commensuration
at high in-plane fields. The similarities between the novel
first-order transition of the $\nu=2$ bilayers and the
commensurate-incommensurate transition of the $\nu=1$ bilayers
help us understand many aspects of our and BM's numerical
findings, such as the confinement of the novel phase transition to
the canted phase and the absence of the new transition in
charge-balanced $\nu=2$ bilayers.

This paper is organized as follows:

In Section \ref{sec:hamilt}, the bilayer Hamiltonian in under the
influence of tilted magnetic fields is presented.

In Section \ref{sec:hf}, the Hartree-Fock procedure, used by Das
Sarma, Sachdev, and Zheng (DSZ) \cite{dassarma:98} to obtain the
phase diagram of the charge-balanced $\nu=2$ bilayer system in
perpendicular field, is extended to the present case of
charge-unbalanced $\nu=2$ bilayer in tilted field.  A similar
procedure was outlined by BM \cite{burkov:02}, who dubbed it the
``unrestricted'' Hartree-Fock approximation.   In
Sec.~\ref{sec:hf}, a detailed presentation of this, unrestricted,
Hartree-Fock procedure is given.

In Section \ref{sec:phdr-0}, the global phase diagram of the
$\nu=2$ bilayers in tilted field is presented.  The phase diagram
of the $\nu =2$ bilayers in perpendicular field has been
previously reported and discussed by several groups of authors
\cite{zheng:97,dassarma:98,brey:99,macdonald:99}.  MRJ
\cite{macdonald:99} in particular, have obtained the full
Hartree-Fock phase diagrams of the $\nu =2$ bilayers for various
combinations of parameters.  MRJ used a very elucidating reduced
Hartree-Fock solution, which elegantly captures the physics of the
$\nu =2$ bilayers.  The only shortcoming of their approximation is
that it does not always give the {\it exact} Hartree-Fock ground
state that is crucial for obtaining the correct collective mode
dispersions from time-dependent Hartree-Fock (as will be presented
in Ref.~\onlinecite{lopatnikova:03b}).  In particular, it gives an
approximate ground state for the canted phase $C$, in which, as
will be shown in Sec.~\ref{sec:phdr-t}, the most interesting
phenomena happen when the magnetic field is tilted.  In addition,
some aspects of the phase diagram, such as the properties of the
zero-tunneling $I$-phase have been essentially uninvestigated.  In
Sec.~\ref{sec:phdr-0} some of these properties are discussed in
detail.

In Section \ref{sec:phdr-t} the main results of this paper are
presented.  In the first part of this section, the physics of the
commensurate-incommensurate transition of the $\nu=1$ bilayers in
tilted fields is reviewed.  In the next part, the possibility of a
similar transition in $\nu=2$ is discussed.  Next, the
quantitative Hartree-Fock results are presented and explained.
The results are summarized in Figs.~\ref{fig:tilt}, \ref{fig:qs},
and \ref{fig:op}.  Figure~\ref{fig:qs} illustrates the
commensuration of spin with the in-plane field at large tilt
angles.  Figure~\ref{fig:tilt} presents a set of global phase
diagrams of the $\nu=2$ bilayers in tilted magnetic fields, which
show the emergence and the evolution of the novel phase
transition, induced by the in-plane field.  Figure~\ref{fig:op}
illustrates the close relationship between the novel
commensurate-commensurate transition (with the involvement of the
spin) and a ``naive'' commensurate-incommensurate transition (in
which the spin is not involved).

Sec.~\ref{sec:concl3} gives a short summary of the results
presented in this paper.



\section{The Bilayer Hamiltonian}
\label{sec:hamilt}

We model the $\nu=2$ bilayer quantum Hall system in tilted
magnetic field by a simple Hamiltonian which includes the five
most important aspects of the system:  Landau level quantization,
Zeeman energy, tunneling between the layers, an external bias
voltage, and the Coulomb interactions between electrons.  Disorder
will be completely neglected throughout this work.

We choose to work in the gauge $\vec{A}(\vec{r}) =
(0,B_{\perp}x,-B_{\parallel} x)$, where $B_{\perp}$ is the
component of the magnetic field perpendicular to the plane of the
sample, and $B_{\parallel}$ is the in-plane component of the field
(the total field is $B_{total} = \sqrt{B_\perp^2 +
B_\parallel^2}$).  If the layers are assumed to be infinitely thin
(an approximation used throughout this work), the
lowest-Landau-level single-electron wavefunctions for each layer
in the gauge $\vec{A}(\vec{r})$ are \be
    \phi_X (\vec{r}) = \frac{1}{\sqrt{l L_y\sqrt{\pi}}}e^{iXy/l^2}e^{-(x-X)^2/2l^2} \label{eq:wavefun},
\ee where $X = -k_y l^2$ are the guiding centers of these
Landau-gauge wavefunctions; $l=\sqrt{\hbar c /e B_{\perp }}$ is
the magnetic length, and $L_y$ is the length of the system in the
$y$-direction.   Throughout this paper, we assume that the
cyclotron energy is much larger than all other energy scales in
the system and restrict our arguments to the lowest Landau level.

In addition to the orbital degrees of freedom, $X$, electrons in
the bilayer systems possess a spin and can be localized in either
layer.  The layer index serves as an additional discrete degree of
freedom --- the isospin --- in bilayer systems.  As a result, the
electron creation operators $c^\dag_{\mu s X}$ are labeled by
three indices:  the orbital index $X$, the spin index $s =
\uparrow$ (or $+1$) and $\downarrow$ (or $-1$), and the layer
index $\mu$ that can take on values $R$ (or $+1$) and $L$ (or
$-1$).   To facilitate the analogy between the spin and the
isospin, we define the electron spin and isospin operators, ${\vec
S}_X$ and ${\vec I}_X$: \bea
    \vec{S}_X & = & \frac{1}{2} \sum_{\mu\, s s'}c^{\dag}_{\mu s\, X}\vec{\sigma}_{ss'}
    c_{\mu s'\, X}^{\phantom{\dagger}} \\
    \vec{I}_X & = & \frac{1}{2} \sum_{s\, \mu\nu}c^{\dag}_{\mu s\, X}\vec{\tau}_{\mu\nu}
    c_{\nu s\, X}^{\phantom{\dagger}},
\eea where $\vec{\sigma}$ and $\vec{\tau}$ are sets of Pauli
matrices, and the spin operator is defined here in the usual
manner \cite{auerbach}.  The total spin and isospin of the system
are defined by the operators ${\cal O}=\sum_{X}{\cal O} _X$, where
${\cal O}=I,S$.

The term in the Hamiltonian representing the Zeeman energy is
simply \be
    H_{Z} = -\sum_X \Delta_Z S_X^z = -\frac{1}{2} \Delta_Z
    (N_{\uparrow} - N_{\downarrow})
\ee where $\Delta_Z = g \mu_B B_{total}$ is the Zeeman splitting,
and $N_\downarrow$ and $N_\uparrow$ are the total numbers of down
and up spins in the system.  Very similarly, the term representing
the bias voltage (i.e.\ the difference in electrostatic potential
between the layers) is written as \be
    H_{V} = -\sum_X \Delta_V I_X^z  = -\frac{1}{2} \Delta_V (N_R -
    N_L),
\ee where $\Delta_V$ is the potential difference between the
layers, and $N_L$ and $N_R$ are the total number of particles in
the left and right wells respectively.  The bias voltage clearly
acts as an effective external field that couples to the isospin.

Within the tight-binding approximation \cite{hu:92}, tunneling
between the layers can also be expressed as an effective external
field coupling to the isospin.  When the magnetic field applied to
the system is perpendicular to it (so that the electrons tunnel
along the direction of the magnetic field and never see any
magnetic flux), the tunneling term can be written as \bea
    H_T^{B_\parallel=0} &=& -\Delta_{SAS}^0 \sum_X I^x_X  \\ &=&
     -\frac{\Delta_{SAS}^0}{2} \sum_{X,s}  \left( c^\dag_{R s X} c_{L s X} +
c^\dag_{L s X} c_{R s X} \right), \eea i.e.\ it acts as an
external field acting on the isospin in the $I^x$-direction. The
coefficient $\Delta^0_{SAS}$ is the symmetric-antisymmetric gap
induced by the tunneling in the absence of the other external
fields and interactions.

When the magnetic field is tilted with respect to the normal to
the sample, the electrons tunneling between the layers pick up an
Aharonov-Bohm phase $[\frac{e}{c} \int {\bf A \cdot dl}] $.  As a
result, the tunneling term acquires phase factors: \bea
   H_T &=&  -\Delta_{SAS} \sum_{X,s } [e^{iQ_{||}X}c^{\dag}_{R s X}c_{L s X}+ \nonumber \\
    & & \mbox{\hspace{65pt}}+e^{-iQ_{||}X}c^\dag_{L s X} c_{R s X}
   ]\\ &=& -\frac{\Delta_{SAS}}{2} \sum_X \left[ e^{iQ_{||}X}I^+_X +
   e^{-iQ_{||}X}I^-_X  \right]  \label{eq:htun}
\eea where the operators $I^{\pm}=I^x\pm iI^y$ are the isospin
raising and lowering operators.  The wavevector $\vec{Q}_{||}$ is
defined as \be \vec{Q}_{||} = \frac{\hat{z}\times
\vec{B}_{||}}{(B_{\perp }l^2/d)} \ee with $d$ the distance between
the two layers \cite{demler:01} (and the in-plane magnetic field,
in our gauge, pointing in the $\hat{y}$-direction).  Note that
$|\vec Q_{||}|$ is proportional to $\tan\theta = B_{||}/B_{\perp
}$, where $\theta $ is the tilt angle.   Interference between the
electrons tunneling in the presence of an in-plane field results
in the reduction of the effective symmetric antisymmetric gap
\cite{hu:92} \be \Delta_{SAS}=\Delta^0_{SAS}e^{-Q_{||}^2l^2/4}.
\label{eq:SASdecay} \ee We note that, had we chosen a different
gauge, the Gaussian decay of $\Delta_{SAS}$ would have remained
although the form of this term would have been different, and the
phase factors could disappear from the tunneling term and reappear
in other terms in the Hamiltonian (see Sec.~\ref{subsec:nu1}).

The three terms of the Hamiltonian ($H_Z$, $H_V$ and $H_t$) given
above all describe single electrons and comprise the
non-interacting part of the Hamiltonian \bea
    H_0 &=& H_Z + H_V + H_T \nonumber \\
    &=& - \sum_{X}\mbox{\Large{[}} \Delta_Z S_X^z +  \Delta_V I^z_X + \nonumber \\
        &\mbox{}& + \frac{\Delta_{SAS}}{2} (e^{iQ_{||}X}I^+_X + e^{-iQ_{||}X}I^-_X) \mbox{\Large{]}} \label{eq:ht0}.
\eea The Coulomb interactions between the electrons are taken into
account by an additional term \bea
    &&H_I =\frac{1}{2\Omega }\sum_{\begin{array}{c} \mbox{\small$X_1X_2$} \\ \mbox{\small$\nu_1,\nu_2$} \\ \mbox{\small$\sigma_1,\sigma_2$} \end{array}}\sum_{\bf q}e^{iq_x(X_1-X_2)}e^{-q^2l^2/2}V_{\mu_1\mu_2}(q)\times \nonumber\\
    &&\mbox{\hspace*{20pt}}\times c^{\dag}_{\mu_1\sigma_1\, X_1+q_yl^2}
    c^{\dag}_{\mu_2\sigma_2\, X_2}
    c_{\mu_2\sigma_2\, X_2+q_yl^2}^{\phantom{\dagger}}
    c_{\mu_1\sigma_1\, X_1}^{\phantom{\dagger}}\label{eq:hi}.
\eea where {\it intra-} and {\it inter}layer Coulomb interactions
are \be
    V_{RR}(q) = \frac{2\pi e^2}{\varepsilon q},\ \ V_{RL}(q) = \frac{2\pi e^2}{\varepsilon q}e^{-dq},
\ee respectively, $d$ is the distance between the layers, and
$\Omega$ is the area of the sample.  The total Hamiltonian is
therefore simply \be
    H = H_0 + H_I.\label{eq:H}
\ee

\section{The Unrestricted Hartree-Fock Approximation}
\label{sec:hf}

\subsection{Trial ground state}
\label{subsec:trialgs}

The Coulomb-interacting Hamiltonian in Eq.~(\ref{eq:H}) is not
tractable exactly, and we solve it using the Hartree-Fock
approximation.  In the usual manner, we assume that the many-body
ground state, $|G\rangle$, is a Slater determinant of
single-particle states and perform a functional minimization of
the expectation value $\langle G|H|G\rangle $ with respect to
these single-particle states.  Under the assumption of
translational invariance in the $\hat{y}$-direction, the most
general single-particle state is a superposition of all the
combinations of spin and isospin degrees of freedom --- $R\uparrow
$, $R\downarrow$, $L\uparrow$, and $L\downarrow$ --- which can be
described as a normalized 4-spinor \be
    W = (w_{R\uparrow},w_{R\downarrow},w_{L\uparrow},w_{L\downarrow}),\label{eq:w}
\ee Such a normalized complex four dimensional vector transforms
under $U(4)$. To make a $\nu=2$ Slater determinant state, we
occupy each orbital $X$ by two particles \cite{burkov:02} \be
    |G \rangle = \prod_{X} f^{\dag}_{1X}f^{\dag}_{2X}|0\rangle , \label{eq:gt}
\ee where \be
    f_{nX}^\dag = \sum_{\mu \sigma} w^n_{\mu \sigma X} c_{\mu \sigma X}^\dag
\ee creates a particle described by Eq.~(\ref{eq:w}).  The
requirement that the operators $f^{\dag}_{nX}$ obey the fermionic
anticommutation relations \be
    \{ f_{mX},f^{\dag}_{nX'}\} = \delta_{mn}\delta_{XX'}\label{eq:fcomm}
\ee is equivalent to an orthonormality constraint on the $W^n$ \be
    \sum_{\mu \sigma}  (w^n_{\mu \sigma X})^*  w^m_{\mu \sigma X}=\delta_{n m}. \label{eq:orthw}
\ee Each element of $U(4)$ specifies four orthonormal $W^n$
spinors.   When the filling fraction is $\nu=2$, the two states
with lowest mean-field energies (which we label $1$ and $2$
throughout the paper) are occupied\footnote{Note that the state
described by Eq.\ref{eq:gt} has a ``gauge" freedom in its
description, as the occupied states 1 and 2 can be rotated into
each other (by a $U(2)$ operation) without changing the physical
state, and similarly the unoccupied states 3 and 4 can be rotated
into each other (by another $U(2)$ operation) without changing the
state. Thus, two electrons occupying a given point $X$ transform
under $U(4)/[U(2) \times U(2)]$.}.

The unrestricted Hartree-Fock ground state given by
Eq.~(\ref{eq:gt}) is very general.  The freedom provided by the
coefficients $w^n_{\mu \sigma X}$ includes a possibility of
non-uniform states, such as stripe states \cite{demler:01}.
However, in this paper, we restrict our attention to states with
uniform density and uniform spin density, ignoring the possibility
of charge density waves or spin density waves.  Our choice is
supported by the absence of instabilities in the excitation
spectra above the uniform-density states, which are calculated in
Ref.~\onlinecite{lopatnikova:03b}.

In order to have a state of uniform spin density and uniform real
density in each layer, the occupation number $N_{\mu \sigma X}$ of
each state $\mu\sigma X$ \be
    N_{\mu \sigma X} = |w^1_{\mu \sigma X}|^2 + |w^2_{\mu \sigma X}|^2
\ee has to be independent of the position $X$.  Thus, we may
write\footnote{Up to the gauge freedom of rotating states 1 and 2
into each other.} \be
    w^n_{\mu \sigma X} = e^{i \phi_{\mu \sigma}(X)} z^n_{\mu \sigma},
\ee where $z^n_{\mu \sigma}$'s are {\it independent} of position
and the only positional dependence arises in the phases.  In the
case of zero in-plane field ($B_{||}=Q_\parallel=0$), it will be
very easy to see below that the lowest energy solution should have
no positional dependence of the phases, so that $\phi_{\mu
\sigma}(X)$ can be taken to be zero. However, in the more general
case of nonzero in-plane field a nontrivial positional dependence
will be favored.

Following BM \cite{burkov:02}, we use a simple trial form for the
positional phase dependence \be \label{eq:phiX}
    \phi_{\mu \sigma} (X) = Q_{\mu \sigma} X
\ee or, equivalently, \be
     f_{nX} = \sum_{\mu\sigma }
      (z^n_{\mu\sigma})^* e^{-iQ_{\mu\sigma}X}c_{\mu\sigma X}.\label{eq:f}
\ee A ground state (Eq.~(\ref{eq:gt})) with nonzero
$Q_{\mu\sigma}$ possesses spin-isospin-wave order, discussed at
length in Sec.~\ref{subsec:cc}

As was mentioned above, the proposed ground state
(Eq.~(\ref{eq:f}) and (\ref{eq:gt})) is not the most general
Slater determinant (Hartree-Fock) state. Indeed, having made a
specific choice (Eq. (\ref{eq:phiX})) of the positional dependence
of the phase, this trial state is not even the most general state
with uniform density and spin density in each layer.   However,
our analysis of the collective modes\cite{lopatnikova:03b} around
the ground states obtained by the minimization of $\langle
G|H|G\rangle$ indicate the stability of these states against
second-order transitions that cannot be described within the
Hilbert space defined by our ansatz. The possibility of phase
transitions into a soliton-lattice state will be discussed in the
concluding section \ref{sec:concl3}.

\subsection{Hartree-Fock minimization}
\label{subsec:min}

We minimize the expectation value of the Hamiltonian (i.e.\ the
trial ground state energy) with respect to the variational
parameters $z^n_{\mu \sigma}$, and $Q_{\mu \sigma}$.  Since \be
    \langle  c^{\dag}_{\mu s X}c_{\nu s' X} \rangle =  e^{i (Q_{\nu s'}  - Q_{\mu s}) X}\sum_{n=1,2} (z^n_{\mu s })^*z^n_{\nu s'},
\ee the expectation value of the ground state energy per unit flux
in our non-uniform ansatz is \bea
    \frac{1}{g}\lefteqn{\langle G|H|G \rangle = } \\
    & & \mbox{} - \frac{1}{2}\sum_{\mu\nu s s'}\left\{ \Delta_Z \delta_{\mu\nu}\sigma^z_{ss'}+ \Delta_V  \delta_{ss'}\tau^z_{\mu\nu} +\right. \nonumber \\
    & & \mbox{} +\Delta_{SAS}\delta_{ss'}[\cos ((Q_{||}-Q_{R s}+Q_{L s})X)\, \tau^x_{\mu\nu} +\nonumber \\
    & & \mbox{} + \left. \sin ((Q_{||}-Q_{R s}+Q_{Ls})X) \, \tau^y_{\mu\nu}]{ }^{ }\right\}\sum_{n=1,2}(z^n_{\mu s })^*z^n_{\nu s'} +\nonumber \\
    & & \mbox{} + H_-\sum_{\mu s }\sum_{n=1,2}|z^n_{\mu s}|^2 [\sum_{s',m=1,2}|z^m_{\mu s'}|^2-1] - \nonumber \\
    & & \mbox{} - \frac{1}{2}\sum_{\mu\nu,ss'}F_{\mu\nu} (-(Q_{\mu s}-Q_{\nu s'})\hat{q}_x)  \times \nonumber \\
    & & \mbox{}\times \sum_{m,n=1,2}(z^n_{\mu s})^*z^n_{\nu s'}(z^m_{\nu s'})^*z^m_{\mu s} \label{eq:gsent} \nonumber,
\eea where $g$ is the Landau level degeneracy, $g=\Omega/2\pi
l^2$, and \bea
    F_{\mu\nu} ({\bf q}) & = & \int\frac{d^2k}{(2\pi)^2}e^{-k^2l^2/2}V_{\mu\nu}(k) e^{i{\bf q\wedge k} l^2} = \nonumber \\
    & = &\int \frac{dk}{2\pi }e^{-k^2l^2/2} V_{\mu\nu}(k)\, k\, J_0(kql^2) \label{eq:fmunu} \\
    H_{-} ({\bf q})  & = &  \frac{1}{4\pi l^2} (V_{RR}({\bf q})-V_{RL}({\bf q})) = \frac{e^2}{\varepsilon l} \frac{1-e^{-dq}}{2ql},\label{eq:hm}
\eea so that $H_- = H_-(0) = \frac{e^2}{\varepsilon l}
\frac{d}{2l}$.  The functions $F_{\mu\nu}$ are monotonically
decreasing functions of $q$ and therefore the higher the
wavevector of the spin-isospin-wave order, the lower the
contribution of exchange to the ground-state energy.  The exchange
term thus favors a uniform state.  The tunneling term, however,
does not contribute to the ground state energy, unless $Q_{R
s}-Q_{Ls} = Q_{||}$, and therefore favors isospin-wave order.  It
is this competition that gives rise to the novel first-order
transition discovered by BM \cite{burkov:02} and is presented in
more detail in the next section.   Using the results of BM
\cite{burkov:02}, we make a simplifying assumption that $Q_{R
\sigma}-Q_{L \sigma} = Q_I$, for both $\sigma = \uparrow$ and
$\downarrow$, and that $Q_{\mu \uparrow}-Q_{\mu\downarrow} = Q_S$,
for $\mu = R$ and $L$. In other words, we reduce the number of
phase parameters from three to two and write \be
    Q_{\mu\sigma} = \frac{\mu}{2}Q_I + \frac{\sigma}{2}Q_S,
\ee where a finite $Q_I$ indicates the presence of an isospin-wave
order, while a finite $Q_S$ reflects the real spin-wave order.

To find a variational minimum of this Hamiltonian, we
differentiate $\frac{1}{g}\langle G|H|G \rangle$ with respect to
$(z^n_{\mu \sigma})^*$ \cite{bethe} under the orthonormality
constraints on $W^n$ (Eq.~(\ref{eq:orthw})).   The resulting set
of minimization conditions can be arranged in the form of a
Schr\"{o}dinger equation: \be
    MZ^n = \epsilon_nZ^n \label{eq:se}
\ee where
$Z^n=(z_{R\uparrow},z_{R\downarrow},z_{L\uparrow},z_{L\downarrow})$,
and $M$ is a $4\times 4$ matrix, which is just the mean-field
single-particle Hartree-Fock Hamiltonian: \bea
    \lefteqn{M_{\nu s';\mu s} = - \Delta_Z \delta_{\mu\nu}\sigma^z_{ss'}- \Delta_V  \delta_{ss'}\tau^z_{\mu\nu} -} \nonumber \\
    && \mbox{} -\Delta_{SAS}\delta_{ss'}\mbox{\Large{[}} \frac{1}{g}\sum_X \cos ((Q_{||}-Q_I)X)\, \tau^x_{\mu\nu} + \nonumber \\
    && \mbox{}+ \frac{1}{g}\sum_X \sin ((Q_{||}-Q_I)X) \, \tau^y_{\mu\nu}\mbox{\Large{]}} +\nonumber \\
    && \mbox{} + 2 H_-\sum_{\mu s }\delta_{\mu\nu}\delta_{ss'}[\sum_{s',m=1,2}|z^m_{\mu s'}|^2-1] - \nonumber \\
    && \mbox{} - F_{\mu\nu} (-[Q_I/2\,(\mu-\nu)+Q_S/2\,(s-s')]\hat{q}_x) \times \nonumber \\
    && \mbox{} \times \sum_{n=1,2}(z^n_{\nu s'})^*z^n_{\mu s} \label{eq:m}.
\eea

The Schr\"{o}dinger equation (Eq.~(\ref{eq:se})) is solved
iteratively \cite{dassarma:98}.  At each iteration, the two
eigenstates corresponding to the lowest eigenvalues are filled
(i.e.\ chosen to be the states 1 and 2).  These lowest-energy
eigenstates $Z^1$ and $Z^2$ are then used to obtain the matrix $M$
for the next iteration.   The procedure is repeated until a
self-consistent solution is achieved.  This solution --- a set of
eigenspinors $Z^n$ sorted according to their eigenvalues ---
defines the lowest energy trial state among the Slater
determinants defined by Eqs.~(\ref{eq:gt}) and (\ref{eq:f})
subject to fixed values of the $Q_{\mu s}$'s. The eigenvalues
$\epsilon_n$ give the binding energy of a particle in the subband
$n$, i.e.\ the energy lost when the particle is taken out of the
system. The sum of individual binding energies does not give the
groundstate energy; the ground state energy is evaluated using
Eq.~(\ref{eq:gsent}). The minimization of the energy of the ground
state over the $Q_{\mu s}$'s is done last \cite{burkov:02}.  Thus,
we find the Hartree-Fock ground state in two steps: First, for
fixed values of $Q_I$ and $Q_S$, we minimize the expectation value
of the ground state energy with respect to $(z^n_{\mu \sigma})^*$.
Then, we minimize the ground state energy with respect to $Q_I$
and $Q_S$.  The results are summarized in the following sections.


\section{Global Phase Diagram of the $\nu=2$ Bilayers in Zero In-Plane Field}
\label{sec:phdr-0}

We start by considering the zero in-plane field case.  As was
mentioned in the introductory section, the phase diagram of the
$\nu=2$ bilayers in perpendicular field has been studied in some
detail \cite{zheng:97,dassarma:98,brey:99,macdonald:99}.  In this
section, we highlight the properties of the zero in-plane field
phase diagram that help elucidate the physics of the $\nu=2$
bilayers in tilted field and discuss a number of issues, to which
little attention has been paid to date.

The Zeeman splitting for a typical bilayer GaAs sample
\cite{pellegrini:98} with filling fraction $\nu=2$ in
perpendicular field is $\Delta_Z^0 \approx 0.01 (e^2/\varepsilon
l)$;  it is set by the properties of the material (the $g$-factor
and the effective electron mass) and the density of the 2DEGs.  We
assume that, in the absence of an external bias voltage, the
density of the 2DEGs in the two layers is the same.  Applying a
finite bias voltage, $\Delta_V$,  perpendicularly to the layers,
one can induce a charge imbalance between the layers.  We assume
that the charge imbalance can be created while keeping both the
filling fraction and the magnetic field constant.  The tunneling
strength, $\Delta^0_{SAS}$ is assumed to be unaffected by the
induced charge imbalance.

It is apparent from Eq.~(\ref{eq:gsent}) that, when the magnetic
field is oriented perpendicularly to the plane of the bilayer
sample, i.e.\ $Q_{||} = 0$, spin- or isospin-wave order is not
favored.  Throughout this section, we therefore assume that $Q_I =
Q_S = 0$.

\begin{figure}[t!]
    \centering
    \rotatebox{270}{\scalebox{0.85}{\includegraphics{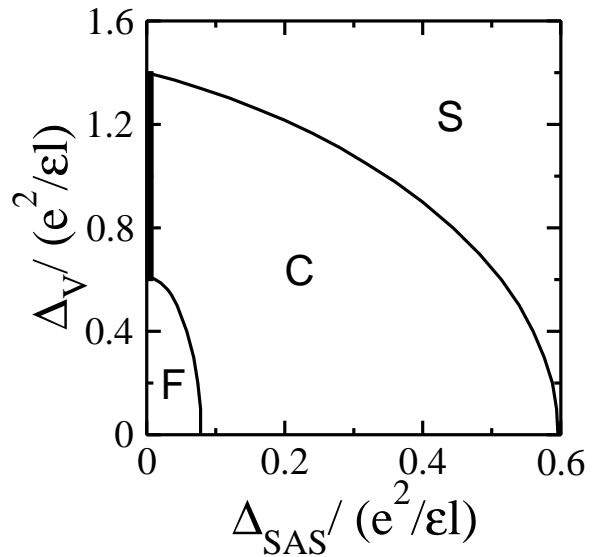}}}
    \caption[Global phase diagram for a $\nu=2$ bilayer sample
    in perpendicular magnetic field, $\Delta_Z^0 = 0.01 (e^2/\varepsilon l)$,
     $d=l$.]{\label{fig:perp} Global phase diagram for a
     $\nu=2$ bilayer sample in perpendicular magnetic field.
      The Zeeman energy is $\Delta_Z^0 = 0.01 (e^2/\varepsilon l)$, the interlayer
      spacing is $d=l$.
      The tunneling strength $\Delta_{SAS}^0$ and the bias
      voltage $\Delta_V$ are given in units of $(e^2/\varepsilon l)$.
      The phase $S$ is the spin-singlet phase, $F$ -- the ferromagnetic
      phase, and $C$ -- the canted phase.  The thick line along
      $\Delta_{SAS}^0 = 0$ represents the $I$ phase.}
\end{figure}

Under these conditions, we obtain the global phase diagram for the
$\nu=2$ bilayers.  A cross-section of the phase diagram for
$\Delta_Z^0 = 0.01 (e^2/\varepsilon l)$ is presented in
Fig.~\ref{fig:perp} \cite{brey:99}.
The phase diagram exhibits four phases: ferromagnetic $F$,
spin-singlet $S$, canted $C$, and the spin-isospin-entangled
non-canted phase $I$ (represented by the thick line along
$\Delta_{SAS}^0 = 0$).   As expected, when both the bias voltage
$\Delta_V$ and the tunneling strength $\Delta_{SAS}^0$ are small,
the Zeeman energy dominates and gives rise to the ferromagnetic
phase.  In the opposite limit, of large  $\Delta_{SAS}^0$ and
$\Delta_V$, the spin-singlet phase is stabilized.  In the
intermediate regime, the Coulomb interactions give rise to the
canted phase when $\Delta_{SAS}^0 \neq 0$, and to the $I$-phase
when $\Delta_{SAS}^0 = 0$.

The topology of the phase diagram is the same for all other finite
values of $\Delta_Z^0$.   For larger values of $\Delta_Z^0$, the
phase space volume of the ferromagnetic phase increases, and the
spin-singlet phase is shifted to higher values of $\Delta_{SAS}^0$
and $\Delta_V$; the width of the canted phase (slowly) decreases.
For smaller values of $\Delta_Z^0$, the opposite effect takes
place:  the volume of the ferromagnetic phase decreases and the
canted phase becomes wider.   The ferromagnetic phase does not
disappear from the phase diagram until $\Delta_Z^0 =0$, when,
within the Hartree-Fock approximation, a many-body phase (the
$\Delta_Z^0 \to 0$ limit of the canted phase) fills the
low-$\Delta_{SAS}^0$-$\Delta_V$ region completely.\footnote{In
exact diagonalization studies of charge-balanced $\nu=2$ bilayer
systems, the spin-singlet phase extends down to very low
$\Delta_{SAS}^0$, when $\Delta_Z^0 =0$.}  In real bilayer samples
$\Delta_Z^0$ is always finite, while $0<\Delta_{SAS}^0 \lesssim
0.08 (e^2/\varepsilon l)$.  As was proposed by Brey, Demler, and
Das Sarma \cite{brey:99}, and is clear form Fig.~\ref{fig:perp},
by sweeping the external bias voltage one can probe the three
phases of the $\nu=2$ bilayer system that occur in the presence of
finite tunneling -- ferromagnetic, canted, and spin-singlet.  In
tilted fields, the $I$-phase can also be attained in the limit of
a large tilt angle (see Sec.~\ref{subsec:cc}).

\subsection{Ferromagnetic phase}
\label{subsec:ferro}

The simplest phase in the phase diagram is the ferromagnetic
phase.  The ferromagnetic ground state has the simple form
$|F\rangle = \prod_X c^{\dag}_{R\uparrow X}c^{\dag}_{L\uparrow
X}|0\rangle$.  It is effectively a single-particle state, which
could occur in the absence of interactions.  The $\nu=2$ system in
the ferromagnetic state can be viewed as two decoupled
spin-polarized $\nu=1$ monolayers.  The state is clearly not
interlayer phase coherent, and an in-plane field therefore would
not affect it.

\subsection{Spin-singlet phase}
\label{subsec:sing}

The spin-singlet state is also a single-particle state.  It is
stabilized by large $\Delta_{SAS}^0$ and/or $\Delta_V$.  The
spin-singlet state, $|S\rangle = \prod_X (z_R c^{\dag}_{R\uparrow
X}+z_L c^{\dag}_{L\uparrow X})(z_R  c^{\dag}_{R\downarrow X}+z_L
c^{\dag}_{L\downarrow X})|0\rangle$, is interlayer phase coherent
unless $\Delta_{SAS}^0 = 0$.  When $\Delta_{SAS}^0 = 0$, the
spin-singlet state is simply $|S\rangle = \prod_X
c^{\dag}_{R\uparrow X} c^{\dag}_{R\downarrow X}|0\rangle$ -- a
$\nu=2$ monolayer quantum Hall state.   In the presence of
interlayer tunneling, the $\nu=2$ bilayer system in the
spin-singlet state can be viewed as two oppositely spin-polarized
$\nu=1$ bilayer systems that possess interlayer phase coherence.
The main difference between the $\nu=2$ bilayer system in the
spin-singlet state and a set of two $\nu=1$ bilayer systems is
that in the latter case the interactions play an important role
alongside tunneling in creating the interlayer phase coherence; in
the spin-singlet state of the $\nu=2$ bilayers interactions are
not important.   The phase space region where the interactions
play an active role in $\nu=2$ bilayers is the region of stability
of the canted phase.  The boundaries of the canted phase therefore
mark the boundaries of the influence of the interactions.
Moreover, unlike in the $\nu=1$ bilayers where tunneling and
exchange work cooperatively, in the $\nu=2$ bilayers tunneling and
exchange are in competition, since the Coulomb interactions favor
the ferromagnetic state (due to their {\it intra-inter}layer
anisotropy).

\subsection{Canted phase}
\label{subsec:c}

\begin{figure}[b]
    \centering
    \rotatebox{270}{\scalebox{0.75}{\includegraphics{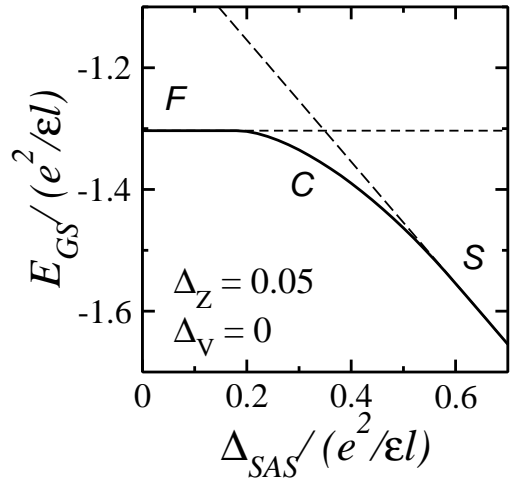}}}
    \caption[Energy profile of charge-balanced $\nu=2$ bilayers with $\Delta_Z =
     0.05 (e^2/\varepsilon l)$ and a range of tunneling amplitudes.]
     {\label{fig:enprof} Energy profile of charge-balanced
      $\nu=2$ bilayers with Zeeman energy $\Delta_Z = 0.05 (e^2/\varepsilon l)$,
      interlayer spacing $d = l$, and a range of tunneling amplitudes.  The dashed line
        represents the energy of the ferromagnetic state.
        The long dashed line is the energy of the spin-singlet state.
        In the absence of interactions, there would be a first order
        phase transition at the intersection of the two dashed lines.
    The interactions effectively ``smooth out'' the profile by
    stabilizing the canted phase.  }
\end{figure}

While the ferromagnetic and the spin-singlet phases are
essentially single-particle phases, stabilized by single-particle
fields, the canted phase is a many-body phase stabilized by the
interactions.  As was mentioned in the introduction, in the
absence of the interactions there would be a first-order phase
transition between the spin-singlet and the ferromagnetic phases.
The interactions can lower the energy of the system by creating
the canted phase (Fig.~\ref{fig:enprof}), and the ground state
energy and width of the canted phase depend on the strength of the
interactions\footnote{The ground state energy and the phase space
volume of the canted phase also depend on the distance between the
layers.  The canted phase can exist only within a finite range of
interlayer distances, comparable to the average intralayer
distance between electrons.}.
As it will be shown in Sec.~\ref{sec:phdr-t}, the importance of
the interactions in the canted phase implies that the phase can be
non-trivially affected by the in-plane component of the magnetic
field.

In the canted phase, the interactions effectively mix the
ferromagnetic and the spin-singlet states \cite{demler:99} giving
rise to a finite magnetization $\langle S^z \rangle$ and
antiferromagnetic spin correlations $|\langle \hat{S}_R^z \times
\vec{S}_R \rangle -\langle \hat{S}_L^z \times \vec{S}_L \rangle
|\neq \vec{0}$, where $\hat{S}_{\mu}^z$ is the unit vector in the
spin-up direction in the layer $\mu$ \cite{dassarma:98}. (One
often defines an antiferromagnetic order parameter ${\cal O}_{xz}$
\cite{macdonald:99,burkov:02}, where ${\cal O}_{\alpha\beta} =
\langle S^{\alpha} \otimes I^{\beta} \rangle$, which is finite
only in the canted phase).  The $U(1)$ symmetry associated with
rotations around $S^z$ is spontaneously broken.  The canted ground
state, $|C\rangle $, has the most general form, Eq.~(\ref{eq:gt}),
and is {\it spin-isospin entangled} --- i.e.\@ it cannot be
decoupled into two independent spin or isospin channels like the
ferromagnetic or the spin-singlet ground states.  The canted
ground state is therefore interlayer phase coherent, which is
confirmed by a non-zero value of $\langle(I^x)^2+(I^y)^2\rangle$.

\subsection{$I$-phase}
\label{subsec:i}

While the canted phase has attracted the most attention
\cite{zheng:97,dassarma:98,demler:99,brey:99,macdonald:99}, the
many-body phase that occurs in charge-unbalanced systems in the
absence of interlayer tunneling is no less interesting. In the
absence of tunneling, the interactions give rise to a {\it
spontaneously} interlayer phase coherent $I$-state
\cite{brey:99,macdonald:99} (with the antiferromagnetic order
parameter, ${\cal O}_{zx}=0$).  The interlayer phase coherence of
the $I$-state is spontaneous since the single-particle fields
$\Delta_Z^0$ and $\Delta_V$ do not stabilize interlayer phase
coherent states.  The $I$-ground state has a simple form
$|I\rangle = \prod_X c^{\dag}_{R\uparrow X}
(z_{R\downarrow}c^{\dag}_{R\downarrow X}
+z_{L\uparrow}c^{\dag}_{L\uparrow X})|0\rangle $, which manifests
the interlayer phase coherence of the state, its spin-isospin
entanglement, and the $U(1)$ spontaneous symmetry breaking. Simply
speaking, the $U(1)$ symmetry of the $I$-state is the freedom to
choose the relative phase of $z_{R\downarrow}$ and $z_{L\uparrow}$
in the expression for $|I\rangle $.  One can also formally
consider the $U(1)$ symmetry associated with rotation around $I^z$
or around $S^z$ (applying $e^{i\theta I^z}$ or $e^{i\theta S^z}$
clearly gives the desired effect).  It is important to note that
the state $|I\rangle$ doe not break the $U(1)\times U(1)$ symmetry
of spin and isospin rotations completely
\cite{dassarma:98,macdonald:99}. This state is an eigenstate of
$I^z+S^z$ with an eigenvalue $g$ (the degeneracy of the Landau
Level), but it mixes states with different $I^z-S^z$ quantum
numbers. So the state $|I\rangle$ breaks the $U(1)\times U(1)$
symmetry down to the diagonal $U(1)$.

The feature of the state $|I\rangle$ to be an eigenstate of $I^z+S^z$
will prove useful in the following discussion of the many-body phases
in the presense of an in-plane field.  In fact, all the possible
ground states of the $\nu=2$ bilayers in the absence of tunneling are
eigenstates of $I^z+S^z$.  If the bias voltage and the Zeeman energy
are positive, all the possible zero-tunneling ground states are
(locally --- for a given orbital quantum number $X$) linear
combinations of a $S^z_X= +1$ spin triplet and an $I^z_X= +1$ isospin
triplet (the ferromagnetic and the spin-singlet states being the two
extremes with only the spin triplet or isospin triplet contributing,
respectively).  In this case, all the zero-tunneling ground states are
eigenstates of $I^z+S^z$ with an eigenvalue $g$, where $g$ is the
Landau level degeneracy.

Even though $I^z+S^z$ is a formal construction, whose expectation
value cannot be directly measured, it can help advance our
understanding of the physics of the $I$-phase and of the canted
phase in the presence of small interlayer tunneling.  Thus, since
applying a rotation $e^{i\theta (I^z+S^z)}$ to the state
$|I\rangle $ generates a trivial phase factor, $I^z+S^z$ can be
treated as the effective direction in which the $I$-state
``points''. In contrast, if tunneling is added to the system,
$I^z+S^z$ does not commute with the Hamiltonian.  The operator
$e^{i\theta (I^z+S^z)}$ then generates a non-trivial rotation of
the canted state around $I^z+S^z$.  When interlayer tunneling is
small, the canted ground state has a large overlap with an
$I$-state, and can be visualized in its Hilbert space as pointing
slightly away from the $I^z+S^z$ direction.  The relevance of this
picture to the $\nu=2$ bilayer physics will become more clear in
the next section (Sec.~\ref{subsec:cc}).

The relation between the canted phase and the $I$-phase is
reminiscent of the relation between phases of the $\nu=1$ bilayers
in the presence and absence of tunneling.  The main similarity is
that, when there is no tunneling between the layers, both the
$\nu=1$ and the $\nu=2$ bilayers support a spontaneously
interlayer phase coherent phase.  There are, however, marked
differences.  The main difference is that the symmetry properties
of the canted phase and the $I$-phase are the same, while in the
$\nu=1$ bilayers spontaneous symmetry-breaking occurs only in the
absence of tunneling.  Moreover, unlike the zero-tunneling phase
of the $\nu=1$ bilayers, the $I$-phase is spin-isospin entangled.
As we will show in the next section, because of their similar
nature, both the $\nu=1$ and $\nu=2$ systems undergo essentially
similar phase transitions in tilted magnetic field;  the systems'
differences, however, lead to surprisingly different behavior of
the $\nu=1$ and $\nu=2$ bilayers in the vicinity of the phase
transitions.


\section{Global Phase Diagram of the $\nu=2$ Bilayers in Tilted Field}
\label{sec:phdr-t}

As was discussed in the previous section, the physics of the
$\nu=2$ bilayers is very rich.  The $\nu=2$ bilayers exhibit a
host of many-body phenomena, such as spontaneous symmetry
breaking, spontaneous interlayer phase coherence, and spin-isospin
entanglement.  To further explore the many-body nature of the
phases and phase transitions of the $\nu=2$ bilayer system, in
this section we study the behavior of the system in the presence
of a tilted magnetic field.

Tilted magnetic fields have been successfully used to study the
role of both spin and layer degrees of freedom in quantum Hall
physics.   In thin monolayer systems, the tilted field technique
was used to investigate the spin-unpolarized fractional quantum
Hall ground states.  The technique is based on the fact that, in
the infinitely thin limit, the orbital motion in a 2DEG depends
solely on the perpendicular component of the magnetic field,
$B_{\perp}$, while the Zeeman energy is proportional to the total
field, $B$.  The Zeeman energy can therefore be increased
independently of the effective interactions in the quantum Hall
monolayer by adding an in-plane field $B_{||}$.  In bilayer
systems, the presence of an in-plane field affects not only the
Zeeman energy but also the {\it relative} orbital motion in the
two layers and, therefore, the tunneling between the layers and
interlayer interactions.

\subsection{Commensurate-incommensurate transition in the $\nu=1$ bilayers: An overview}
\label{subsec:nu1}

The coupling of the tilted field to the layer degree of freedom is
easier to demonstrate using the thoroughly studied $\nu=1$
bilayers \cite{wen:92,murphy:94,yang:94,moon:95,yang:96} as an
example. In $\nu=1$ bilayers the spin degree of freedom is frozen
out by the ferromagnetic exchange \cite{zheng:97} and, in the
absence of an external bias voltage, the physics of the system is
fully determined by the interplay of tunneling and Coulomb
interactions.  As was discussed above, tunneling
(Eq.~(\ref{eq:htun})) can be considered as an external field that
couples to $I^x$; Coulomb interactions give rise to a charging
energy (from the direct term) and an anisotropic
isospin-stiffness, both of which favor the isospin-$xy$ plane
\cite{yang:94}. When the layers are separated by a distance
comparable to the distance between the electrons in a single
layer, the anisotropic Coulomb interactions support a
spontaneously interlayer phase coherent ground state even in the
absence of tunneling \cite{wen:92}.  The $U(1)$ symmetry
associated with rotations around $I^z$ is spontaneously broken.
Tunneling, always present to some degree in real samples, breaks
this symmetry, but it does not destroy the interlayer phase
coherence.  Instead, in the absence of an in-plane field,
tunneling acts cooperatively with the interactions to stabilize
the interlayer phase coherent state \cite{murphy:94,yang:94}.

A finite in-plane field introduces a competition between tunneling
and the Coulomb interactions in $\nu=1$ bilayers
\cite{yang:94,yang:96}.  In the presence of an in-plane field, the
tunneling term (Eq.~(\ref{eq:htun})) now favors an isospin-wave
ground state, in which the isospin twists around the
$I^z$-direction with the wavevector $Q_{||}$.  Exchange
interactions, on the other hand, favor a uniform configuration.
The competition between tunneling and exchange results in the {\it
commensurate-incommensurate} transition between the isospin-wave
{\it commensurate} and the uniform {\it incommensurate} phases.

The picture of the commensurate-incommensurate transition outlined
above is somewhat simplistic \cite{yang:94,yang:96}. The
transition, in fact, does not happen directly from the
commensurate to the incommensurate state, but instead occurs
through formation of a soliton-lattice phase.  When, as the
in-plane field is increased, the losses in exchange energy due to
twisting become approximately equal to the expectation value of
the tunneling term, the system can optimize its ground state
energy by forming a soliton in an otherwise commensurate state
\cite{yang:94,yang:96,cote:95,hanna:01}. By forming a soliton the
system recovers some exchange energy while saving most of the
tunneling energy.  The commensurate-incommensurate transition by
means of soliton-formation is the Talapov-Pokrovsky
commensurate-incommensurate transition\cite{yang:94}.

The microscopic description of the physics of a bilayer system in
an in-plane field depends on a particular choice of the gauge, but
the underlying picture of an induced competition between tunneling
and exchange is valid for any gauge \cite{yang:96}.  Thus, for
example, in the gauge  $\vec{A} = (0,B_{\perp}x-B_{||}z,0)$, the
tunneling electrons acquire no Aharonov-Bohm phase and the
tunneling term in the microscopic Hamiltonian is unaffected
\cite{hu:92,murphy:94}.  However, in this gauge it is the
interlayer exchange that acquires an additional phase in this
case, so that it is the interaction term that now favors an
isospin-wave.

\subsection{Possibility of commensurate-incommensurate transition in $\nu=2$ bilayers}
\label{subsec:naive}

To create a framework for the interpretation of our numerical
results, we start by a qualitative discussion of the behavior of
the $\nu=2$ bilayers in tilted field.  We consider the theoretical
possibility of the commensurate-incommensurate transition in the
$\nu=2$ bilayers, much like it has been done for the $\nu=1$
bilayers:  We assume for simplicity that the
commensurate-incommensurate transition happens between a
commensurate state that maximizes tunneling at the expense of
exchange (i.e.\ $Q_I = Q_{||}$ and $Q_S = 0$) and an
incommensurate state that maximizes exchange while losing all the
tunneling energy ($Q_I = Q_S = 0$).  This is the scenario that we
called simplistic in our discussion of the $\nu=1$ bilayers since
better mathematical models indicate that the
commensurate-incommensurate transition occurs through a succession
of soliton lattice phases rather than directly.  Nevertheless, the
``naive'' commensurate-incommensurate transition serves as a good
simple approximation that allows one to make experimentally
relevant predictions and explanations without any numerical
calculations. We therefore expect that the naive
commensurate-incommensurate approximation will capture some of the
physics of the $\nu=2$ bilayers in tilted fields as well.

The naive commensurate-incommensurate transition can happen only
in the canted phase, since it is the only phase in which tunneling
and exchange are comparable to each other.   The other two phases
--- the ferromagnetic phase and spin-singlet phase --- essentially
mark the regions in the phase space where interactions are less
important than the single particle fields.  We therefore expect
the spin-singlet phase to be always commensurate.  The
ferromagnetic state is not interlayer phase coherent and therefore
the in-plane field cannot affect it.  The canted phase however can
be commensurate at lower in-plane fields, but it may turn
incommensurate when the in-plane fields are so high that the
losses in exchange become greater than the contribution of the
tunneling term.

The naive commensurate-incommensurate transition however cannot
happen in charge-balanced $\nu=2$ bilayers.  In order for the
commensurate-incommensurate transition to happen, there should
exist an incommensurate state, which would become lower in energy
than the corresponding commensurate state as the tilt angle is
increased.  Since the tunneling contribution to the energy of the
naive incommensurate state is zero, the incommensurate state is
equivalent to the state of the system with no interlayer tunneling
(all other parameters unchanged).   In charge-balanced $\nu=2$
bilayers, the zero-tunneling ground state is always ferromagnetic.
As is clear from Fig.~\ref{fig:enprof}, the energy of a
ferromagnetic state created in a system with no tunneling and
given intralayer interactions is always higher than that of a
canted state created by adding a finite amount of tunneling to
this system (all other things being equal).   Therefore, there can
be no naive commensurate-incommensurate transition in a
charge-balanced $\nu=2$ bilayer system.

The situation changes if a finite bias voltage can be applied to the
system.  As was mentioned in Sec.~\ref{sec:phdr-0}, in the presence of
finite bias voltage one can obtain an interlayer phase coherent phase
--- the $I$-phase --- even in the absence of tunneling.  The ground
state energy of the $I$-state depends not only on the relative
strength of the Zeeman energy and the external bias voltage, but
also on the interlayer Coulomb interactions.  Therefore, it is
straightforward to argue that, in charge-unbalanced samples with
small tunneling amplitudes, a commensurate-incommensurate
transition may be possible: Let us consider such a sample.  In
perpendicular field, the many-body state of the sample is a canted
state.  Because the tunneling amplitude is small, this canted
state is just a slight perturbation of the corresponding
zero-tunneling $I$-state.  The interaction energies of the two
states are very close; the energy of the canted phase is lower
largely due to the tunneling term.  As the magnetic field is
tilted, the isospin starts twisting commensurately with the
in-plane field, thereby losing some interlayer exchange. When the
losses in exchange energy become equal to the tunneling energy,
the energies of the canted commensurate state and the
corresponding zero-tunneling $I$-state become approximately equal.
At slightly higher tilt angles, a transition to the $I$-state
clearly would occur.  In anticipation of this discussion, at the
start of the chapter, we named the zero-tunneling many-body phase
the $I$-phase, in which $I$ stands for ``incommensurate''.

We also can expect that the in-plane field will not only induce a
new transition, but also affect the location of the second-order
phase transitions between the spin-singlet, the canted and the
ferromagnetic phases in the phase space.  The reduced effective
interlayer interactions in the commensurate states will
destabilize the canted commensurate phase, and its two phase
boundaries will move toward each other as the in-plane field is
increased.  The relative position of the boundaries of the
incommensurate phase will be nearly constant with the increasing
tilt angle.   At larger in-plane fields another effect, which we
so far have have ignored, will become important --- the dependence
of the Zeeman energy and the tunneling amplitude on the in-plane
field.   The increasing Zeeman energy and the decreasing tunneling
will eventually drive the system into a ferromagnetic state as the
tilt angle becomes very large.  (At very large in-plane fields,
however, our infinitely-thin layer approximation loses its
validity completely, and therefore this discussion becomes purely
theoretical.)
\begin{figure*}[t]
    \centering
    \rotatebox{270}{\scalebox{0.7}{\includegraphics{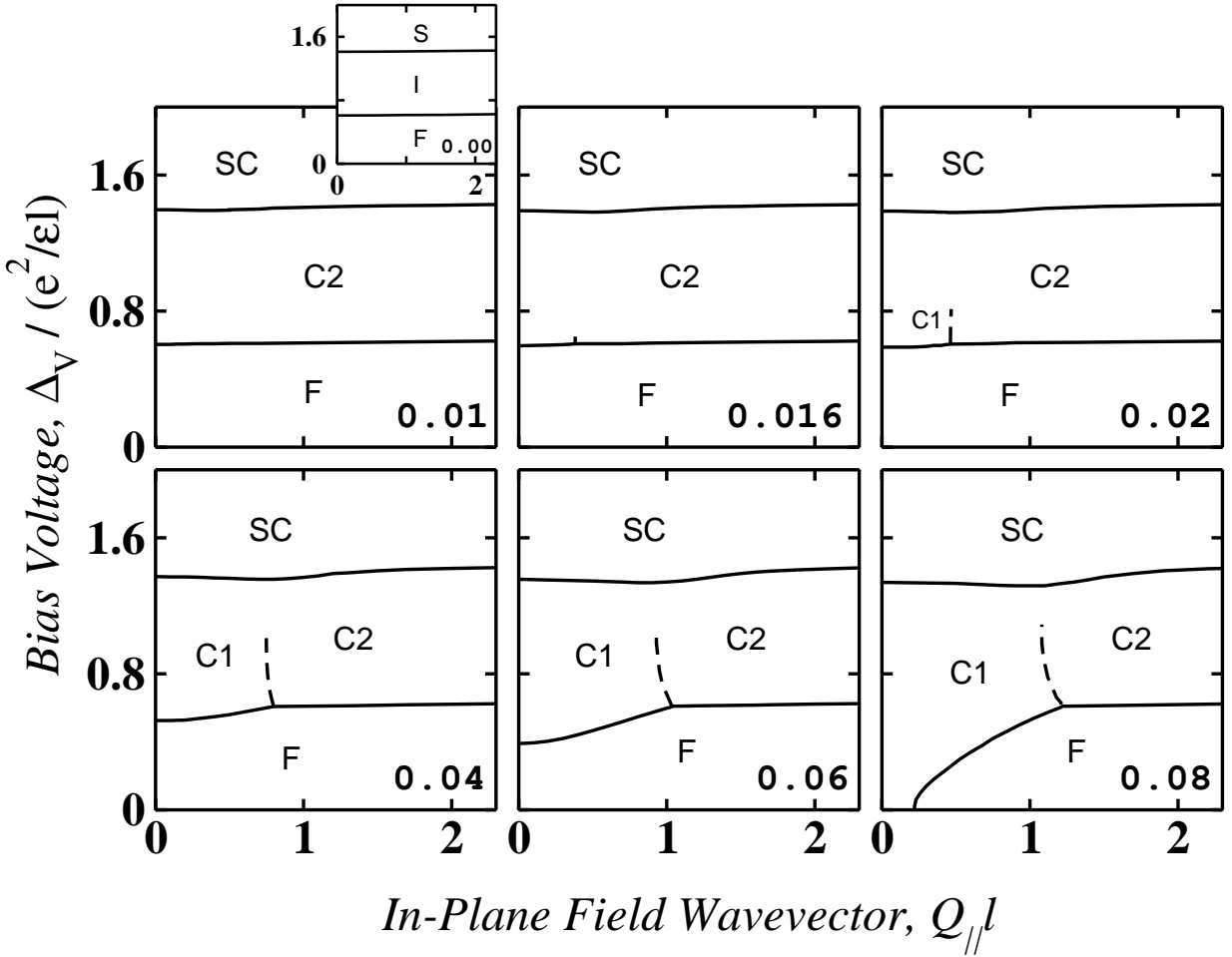}}}
    \caption[Global phase diagrams for $\nu=2$ bilayer samples of
    different tunneling strengths in tilted magnetic field.]
    {\label{fig:tilt} Global phase diagrams for $\nu=2$ bilayer
     samples of different tunneling strengths in tilted magnetic
     field.  The tunneling amplitudes, $\Delta_{SAS}^0$ are given
     in the lower right corner of each panel. The Zeeman energy,
      $\Delta_Z^0 = 0.01 (e^2/\varepsilon l)$, and the distance
      between the layers $d=l$, are the same for all the phase diagrams.
       The solid lines indicate second-order phase transitions, the
       dashed lines -- first-order.  The phases are:  $S$ - spin-singlet,
       $F$ - ferromagnetic, $I$ - incommensurate, $SC$ - spin-singlet
        commensurate, $C1$ - canted commensurate, $C2$ - canted spin-isospin
        commensurate.}
\end{figure*}

In the next subsection we explain that the commensurate-incommensurate
transition at $\nu=2$ is sufficiently different from the simple picture
presented above. Hence, the arguments presented here should be considered
only as a motivation for a more detailed discussion in subsecuent
sections.

\subsection{Global phase diagram of $\nu=2$ bilayers in tilted magnetic field}
\label{subsec:phdr6}

The Hartree-Fock phase diagrams are presented in
Fig.~\ref{fig:tilt}.   The axes on the phase diagrams are the bias
voltage, $\Delta_V$, and the in-plane field wavevector $Q_{||}$.
We find this choice of axes convenient, since current experimental
techniques allow to tune both the bias voltage and the in-plane
field {\it in situ} across a wide range.  The other parameters of
a bilayer sample --- the perpendicular-field Zeeman splitting,
$\Delta_Z^0$, the perpendicular-field tunneling amplitude,
$\Delta_{SAS}^0$, and the distance between the layers, $d$ ---
are, to a good approximation, intrinsic to a given sample.  For
each phase diagram, we fix these parameters at values typical of
real samples: $\Delta_Z^0 = 0.01 (e^2/\varepsilon l)$,
$\Delta_{SAS}^0 \leq 0.08 (e^2/\varepsilon l)$, and $d=l$.   Each
phase diagram in Fig.~\ref{fig:tilt} therefore corresponds to a
single sample with $\Delta_Z^0 = 0.01 (e^2/\varepsilon l)$, $d=l$,
and the value of $\Delta_{SAS}^0$ given in the lower right corner
of the phase diagram.

The unrestricted  Hartree-Fock calculation does provide evidence
for a new phase transition \cite{burkov:02} (the dashed line in
Fig.~\ref{fig:tilt}), which possesses the properties we
qualitatively predicted for the naive commensurate-incommensurate
transition:  The first-order transition occurs only within the
canted phase and only in the presence of a finite bias voltage.
Moreover, the canted commensurate phase shrinks as the in-plane
field is increased, but the decrease in the width of the phase
stops after the first-order transition.   However, instead of the
incommensurate phase, we find an interesting, what seems like a
doubly-commensurate phase, in which both the isospin and the spin
components are commensurate with the in-plane field
\cite{burkov:02}.  That is to say, throughout the interlayer phase
coherent region --- in the phases $SC$, $C1$ and $C2$ on
Fig.~\ref{fig:tilt} --- the wavevector of the isospin-wave $Q_{I}
= Q_{||}$.
\begin{figure}
    \centering
    \rotatebox{270}{\scalebox{0.75}{\includegraphics{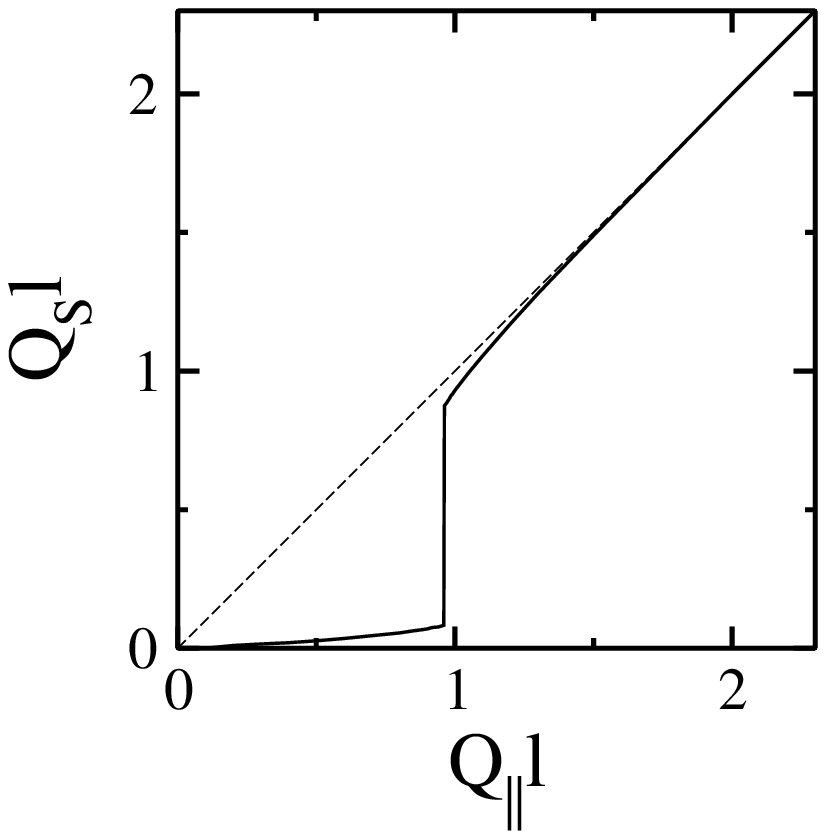}}}
    \caption[Evolution of the spin-wave wavevector $Q_S$ as a function of
     the in-plane field wavevector $Q_{||}$.]{\label{fig:qs}
      Evolution of the spin-wave wavevector $Q_S$ (solid line)
      as a function of the in-plane field wavevector $Q_{||}$
      (given in units of $1/l$). The dashed line is $Q_I = Q_{||}$,
       given for comparison.  The wavevector $Q_S$ is small in
       the phase $C1$ (low $Q_{||}$) and abruptly jumps to
       $Q_S \approx Q_{||}$ at the phase transition to $C2$.
        The Zeeman energy in this figure is $\Delta^0_Z = 0.01(e^2/\varepsilon l)$,
        the interlayer spacing is $d =l$, the tunneling constant is $\Delta_{SAS}^0 = 0.06 (e^2/\varepsilon l)$,
        and the external bias voltage is $\Delta_V = 0.8 (e^2/\varepsilon l)$.}
\end{figure}
In the canted phases $C1$ and $C2$ however the spin-wave
wavevector is also nonzero.  It is almost zero in the $C1$ phase,
except near the phase transition boundary, but it is close to
$Q_{S} = Q_{||}$ in the $C2$ phase (Fig.~\ref{fig:qs} and
Ref.~\onlinecite{burkov:02}).  The phase transition between the
two canted phases is first-order, terminating at a critical point.
The onset of the first-order transition occurs at a critical
tunneling amplitude $\Delta_{SAS}^0 \approx 0.015 (e^2/\varepsilon
l)$.  As the tunneling amplitude $\Delta_{SAS}^0$ is increased,
the $C1$--$C2$ transition becomes more prominent and a higher
in-plane field is needed to induce it.  The presence of the
in-plane component of the magnetic field thus leads to a phase
transition that is clearly related to the
commensurate-incommensurate transition, but possesses some
unexpected properties that invite a physical explanation.

We emphasize that a simple picture of the commensurate-incommensurate
transtion should not be carried directly from $\nu=1$ to $\nu=2$.  In the
former case it appears as a result of the competition between the
single particle tunneling energy and the exchange part of the Coulomb
interaction. At $\nu=2$ we have $Q_I=Q_\parallel$ in all of the canted
phase (both C1 and C2), which optimizes the tunneling term. The origin
of the C1-C2 transtions is the competition of the exchange terms in
equation (31). The wavevectors of the exchange terms in this equation
are given by $Q_s$, $Q_\parallel-Q_s$, and $Q_\parallel+Q_s$ (we used
$Q_I=Q_\parallel$). So different exchange terms would be minimized for
different values of $Q_s$. It is useful to consider the variational
energy of the ground state in (31) as a function of $Q_s$ for
different points in the phase diagram, $E(Q_s)$. When we start near
the base of the first order transition and far from the critical point
(see Fig 3), the function has two local minima: one for $Q_s$ close to
zero and the other for $Q_s$ close to $Q_\parallel$.  When we are
on the C1 side of the transition, the former is the global minimum,
and when we are on the C2 side, the latter corresponds to the true
ground state. As we move along the first order line toward the
critical point (by increasing the gate voltage), the positions of the
two local minima are moving together until at the critical point the
merge into a single minimum. Anywhere above the critical point the
system has only one local minimum in $E(Q_s)$. It is also useful to
point out that there is a region of metastability of C1 and C2 phases
around the first order line separating them. We expect interesting
hysteresis effects  to occur in this region.

\subsection{Canted commensurate phases in $\nu=2$ bilayers}
\label{subsec:cc}

\begin{figure}
    \centering
    \rotatebox{270}{\scalebox{0.75}{\includegraphics{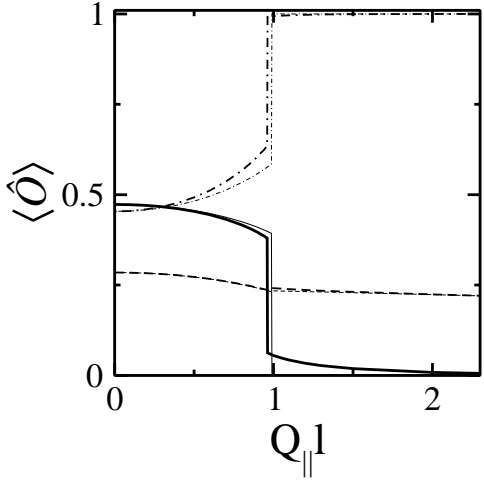}}}
    \caption[Comparison of the evolution of various order parameters
     across the $C1$--$C2$ and naive commensurate-incommensurate phase
     transitions.]{\label{fig:op} Expectation values  $\langle I^x_Q \rangle$
      (solid lines), $\langle I^z \rangle$ (dashed lines),
      and $\langle I^z+S^z \rangle$ (dash-dotted lines),
       per flux quantum, across the $C1$--$C2$ (thick lines)
       and the naive commensurate-incommensurate (thin lines)
       phase transitions. The Zeeman energy is $\Delta^0_Z = 0.01(e^2/\varepsilon l)$,
        the interlayer spacing is $d=l$, the tunneling constant is $\Delta_{SAS}^0 = 0.06 (e^2/\varepsilon l)$,
        and the external bias voltage is $\Delta_V = 0.8 (e^2/\varepsilon l)$. }
\end{figure}

The most surprising part of the phase diagram is the $C2$ phase,
in which both the spin and the isospin degrees of freedom are
commensurate with the in-plane field \cite{burkov:02}.  The
location of the $C2$ phase on the phase diagram, so similar to
that expected of the incommensurate phase, strongly suggests that,
despite its apparent complexity, the $C2$ phase is simply related
to the naive incommensurate phase.  The close relationship between
the $C2$ phase and the $I$-phase becomes more clear if one
considers the expectation values of spin and isospin operators.
In Fig.~\ref{fig:op}, we plot three expectation values: $\langle
I^x_Q \rangle = \sum_X \langle e^{iQ_{||}X}I^+_X +
e^{-iQ_{||}X}I^-_X \rangle$, $\langle I^z \rangle$, $\langle
I^z+S^z \rangle$ for a sample with $\Delta_Z^0 = 0.01
(e^2/\varepsilon l)$, $\Delta_{SAS}^0 = 0.06 (e^2/\varepsilon l)$,
and the bias voltage held at $\Delta_V = 0.8 (e^2/\varepsilon l)$
as we move across the first-order transition by increasing
$Q_{||}$.
The operator $I^x_Q$ is the tunneling operator in the Hamiltonian
of the system, and therefore its expectation value indicates if
tunneling contributes to the energy of a particular state.  The
expectation value  $\langle I^x_Q \rangle$ is zero in the naive
incommensurate state.  The naive  incommensurate state, the
$I$-state, is an eigenstate  of the operator $I^z+S^z$ with the
eigenvalue $g$.  Thus,  $\frac{1}{g}\langle I^z+S^z \rangle =1$ in
this state. To distinguish the $I$-state from the ferromagnetic
state and the fully charge-unbalanced spin-singlet state, which
are also eigenstates of $I^z+S^z$, we also plot $\langle I^z
\rangle$, which satisfies $0<\frac{1}{g}\langle I^z \rangle <1$ in
the $I$-state.  The thinner lines in Fig.~\ref{fig:op} represent
the expectation values obtained under the assumption that the
system undergoes the naive commensurate-incommensurate transition
(the spin-wave wavevector, $Q_S$, is held at 0).   The
discontinuity in the expectation values marks the
commensurate-incommensurate transition.  As expected, after the
transition into the incommensurate phase, $\frac{1}{g}\langle
I^x_Q \rangle = 0$ and $\frac{1}{g}\langle I^z+S^z \rangle =1$.
The thick lines represent the expectation values that we obtain
for the system allowed to undergo the $C1$--$C2$ transition.  The
expectation values obtained from the full unrestricted
Hartree-Fock solution and those obtained under the assumption of
the naive commensurate-incommensurate transition exhibit
strikingly similar behavior.    The main difference is that,
unlike in the $I$-phase, the expectation value of the $I^x_Q$
operator in the $C2$ phase is small but finite.  This means that
there is a contribution from the interlayer tunneling to the
ground state energy in the $C2$ phase.  It is therefore clear that
the $C2$ phase is the optimized version of the $I$-phase much like
the soliton-lattice phase of the $\nu=1$ bilayers is the optimized
$\nu=1$ incommensurate state.

\begin{figure*}[t]
    \centering
    \scalebox{0.7}{\includegraphics{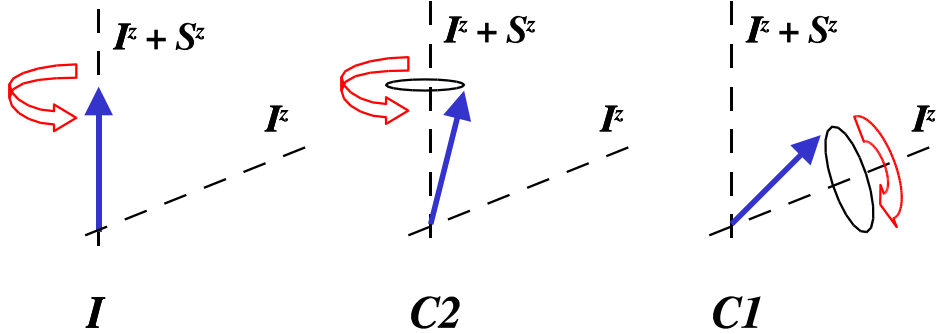}}
    \caption[Schematic representation of the commensurate
    $C1$ and $C2$ states.]{\label{fig:whatis} Schmatic
    representation of the comensurate $C1$ and $C2$ states.
    In the left figure, an $I$-state is represented.
    The $I$-state is shown in Sec.~\ref{subsec:i} to be an
    eigenstate of the operator $I^z+S^z$.  Thus, the $I$ state
    is represented as a vector pointing in the $I^z+S^z$ direction.
    In the presence of tunneling, the system can satisfy the Aharonov-Bohm
    phases in the tunneling term by
    winding around either the $I^z+S^z$ axis, yielding the $C2$-commensurate
    state shown above, or
    by winding around the $I^z$ axis yielding the $C1$-commensurate state.
    The $C2$-state asymptotically
    approaches the $I$-state in large in-plane field where
    the winding becomes very fast
    and therefore must be very tight (due to ``spin stiffness'').}
\end{figure*}

Unlike the $\nu=1$ bilayer system, the $\nu=2$ bilayer system
possesses the spin degrees of freedom which it can use to optimize
its ground state energy around the commensurate-incommensurate
transition.  The system can satisfy the Aharonov-Bohm phases in
the tunneling term by winding either around $I^z$ or $I^z+S^z$.
Winding around $I^z+S^z$ clearly does not affect the Zeeman and
the bias voltage terms and in some circumstances can cost less
exchange energy:  The $I$-state is an eigenstate of the $I^z+S^z$
operator and, as we argued in Sec.~\ref{subsec:i}, is invariant
under rotation around $I^z+S^z$.   If the $I$-state is somehow
perturbed so that it becomes slightly canted, for example because
a small amount of tunneling is present, the state is no longer
invariant under rotations around $I^z+S^z$, but a precession
around $I^z+S^z$ does not cost much exchange energy. (A useful
analogy is a Heisenberg ferromagnet with all spins pointing in the
same direction, $\langle S^z\rangle = +1/2$.  If the spins are
made to tilt away from the positive $S^z$-direction and precess
around it, very little spin-stiffness energy is lost.) As the
in-plane field is increased, the winding around $I^z+S^z$ becomes
faster and tighter.  This causes the phase $C2$ to asymptotically
approach $I$.

The $C1$ and $C2$ phases are both canted and have the same
symmetry properties.  It is therefore not surprising that they are
connected on the phase diagram, i.e.\ they are essentially the
same phase.  The qualitative difference between $C1$ and $C2$ is
in the involvement of the spin degree of freedom in the quenching
of the in-plane magnetic field.  In our mean-field solution the
wavevector of the spin-wave does not always jump between the
qualitatively understood cases of $Q_S = 0$ and $Q_S = Q_{||}$,
but it can change gradually.  The wavevector $Q_S$ changes
gradually when $C1$ turns into $C2$ via a cross-over, at larger
values of $\Delta_V$, when the canted state has a large overlap
with the $I$-phase ($\frac{1}{g}\langle I^z+S^z \rangle $ close to
1).  In fact, when $\Delta^0_{SAS}$ is very small, $\Delta^0_{SAS}
\leq 0.015 (e^2/\varepsilon l)$, the first order transition
disappears altogether, since the canted phase is so close to the
$I$-phase that the canted phase always has a $C2$ flavor to it.


\section{Conclusions}
\label{sec:concl3}

To summarize, we have obtained the global phase diagram of the
$\nu=2$ bilayers in tilted magnetic field (Fig.~\ref{fig:tilt}).
We found that, in {\it charge-unbalanced} $\nu=2$ bilayers, a
finite in-plane component of the magnetic field can induce a
first-order phase transition between two commensurate canted
phases $C1$ and $C2$.  The phase $C1$ possesses isospin-wave
order, commensurate with the in-plane field;  in the phase $C2$
commensurate {\it spin}-wave order is induced alongside with the
isospin-wave order.  Both $C1$ and $C2$ phases spontaneously break
a global $U(1)$ symmetry, and are technically the same phase.
Indeed, in the phase diagrams in Fig.~\ref{fig:tilt}, phases $C1$
and $C2$ are topologically connected, and the first-order
transition between them terminates at a critical end-point.

The physics of the commensurate canted phases was discussed in
detail in this paper.  The behavior of the $\nu=2$ bilayers in
tilted magnetic fields was compared to that of their $\nu=1$
counterparts.  The phase $C1$ was found to be analogous to the
commensurate phase of $\nu=1$ bilayers, while the phase $C2$ was
linked to the incommensurate phase.  As was predicted by MRJ
\cite{macdonald:99}, the $U(1)$-symmetry-broken $I$-phase, which
had been predicted to exist in the absence of tunneling in
charge-unbalanced $\nu=2$ bilayer systems, was found to play the
role of a ``naive'' incommensurate phase in $\nu=2$ bilayers (akin
to the ``naive'' --- translationally invariant --- incommensurate
phase of $\nu=1$ bilayers (see Sec.~\ref{subsec:nu1})).  In this
paper, the $C2$ phase was argued to be an optimization of the
naive incommensurate phase, much like the soliton lattice phase in
$\nu=1$ bilayers in tilted fields is an optimization of the naive
incommensurate phase in this system.

The rapid convergence of the spin-isospin commensurate canted
phase to the $I$-phase can be used to study the $I$-phase.  Much
like the canted phase, the $I$-phase also possesses a number of
intriguing many-body properties.  However, the $I$-phase can occur
only in the absence of interlayer tunneling -- a condition
impossible in a typical bilayer sample.  Tilting the magnetic
field allows one to access the $I$-phase in an experimental
setting and study its properties.

The possibility of the formation of an interim soliton phase
around the $C1$--$C2$ first order phase transition cannot be ruled
out with certainty in our approximation.  However, the energy of a
soliton phase in $\nu=1$ bilayers converges to the energy of the
corresponding naive (translationally invariant) incommensurate
phase \cite{cote:95,hanna:01}, much more rapidly than does the
energy of $C2$ converges to the energy of the corresponding
$I$-state.  We may therefore conclude that, even if the soliton
phase in $\nu=2$ bilayers is possible, it will not occupy a
significant amount of phase space.

In this paper we present a heuristic argument and numerical
evidence that no new transition occurs in {\it charge-balanced}
$\nu=2$ bilayers.  This is indeed consistent with the inelastic
light-scattering results by Pellegrini et al.
\cite{pellegrini:97,pellegrini:98}:  In their experiments,
Pellegrini et al.\ used the tilted-field technique to sweep over a
range of Zeeman energy {\it in situ}.   No perpendicular bias
voltage was applied to the bilayer system;  the maximum tilt angle
was $\theta =45^o$.  Pellegrini et al.\  obtained encouraging
evidence of the existence of the expected phase transitions
between the spin-singlet and the canted phases, as well as between
canted and ferromagnetic phases.  No other transitions have been
reported.

The Hartree-Fock approximation, which we used to obtain our
results, had been shown to be robust for the $\nu=2$ bilayers in
perpendicular fields \cite{macdonald:99,schliemann:00}.  The phase
diagrams obtained in the Hartree-Fock approximation closely match
those obtained using exact diagonalization \cite{schliemann:00}.
While the Hartree-Fock approximation overestimates the size of the
canted region on the spin-singlet side, it reproduces the boundary
between the canted and the ferromagnetic phases apparently {\it
exactly} \cite{schliemann:00}.  Since the novel phase transition
occurs closer to the ferromagnetic side of the canted phase, it is
reasonable to assume that the quantum fluctuations, not taken into
account in the HF approximation, will not wash it out.  The
quantum fluctuations will probably effectively renormalize the
canted phase and make the first order-phase transition terminate
closer to the ferromagnetic-canted line.


\section*{Acknowledgements}

The authors would like toacknowledge helpful conversations with
J.~H.~Cremers, S.~Das Sarma, B.~I.~Halperin, L.~Marinelli,
A.~Pinczuk, D.~Podolsky, L.~Radzihovsky, G.~Refael,
Y.~Tserkovnyak, D.-W.~Wang, and X.-G.~Wen.  This work has been in
part supported by NSF-MRSEC grant No.~DMR-02-13282 and by the NSF
grant No.~DMR-01-32874.  A.L. would like to thank Lucent
Technologies Bell Labs for hospitality and support under the GRPW
program.


\end{document}